\documentclass[twocolumn]{aastex631}
\usepackage{amsmath}
\usepackage{physics}

\usepackage{pythonhighlight}

\begin{document}

\title{Deconvolving X-ray Galaxy Cluster Spectra Using a Recurrent Inference Machine}

\correspondingauthor{Carter Rhea}
\email{carterrhea@astro.umontreal}

\author[0000-0003-2001-1076]{Carter Lee Rhea}
\affiliation{Département de Physique, Université de Montréal, Succ. Centre-Ville, Montréal, Québec, H3C 3J7, Canada}
\affiliation{Centre de recherche en astrophysique du Québec (CRAQ)}
\affiliation{Iuvo-ai, Montréal, Québec, Canada}

\author[0000-0001-7271-7340]{Julie Hlavacek-Larrondo}
\affiliation{Département de Physique, Université de Montréal, Succ. Centre-Ville, Montréal, Québec, H3C 3J7, Canada}

\author[0000-0001-8806-7936]{Alexandre Adam}
\affiliation{Département de Physique, Université de Montréal, Succ. Centre-Ville, Montréal, Québec, H3C 3J7, Canada}
\affiliation{Mila - Quebec Artificial Intelligence Institute, Montreal, Qu\'ebec, Canada}

\author[0000-0002-0765-0511]{Ralph Kraft}
\affiliation{Center for Astrophysics $\vert$ Harvard \& Smithsonian, 60 Garden Street, Cambridge, MA 02138, USA}

\author[0000-0003-0573-7733]{\'Akos Bogd\'an}
\affiliation{Center for Astrophysics $\vert$ Harvard \& Smithsonian, 60 Garden Street, Cambridge, MA 02138, USA}

\author[0000-0003-3544-3939]{Laurence Perreault-Levasseur}
\affiliation{D\'epartement de Physique, Universit\'e de Montr\'eal, Succ. Centre-Ville, Montr\'eal, Qu\'ebec, H3C 3J7, Canada}
\affiliation{Mila - Quebec Artificial Intelligence Institute, Montreal, Qu\'ebec, Canada}
\affiliation{Center for Computational Astrophysics, Flatiron Institute, New York, USA}

\author[0009-0003-0932-2487]{Marine Prunier}
\affiliation{Département de Physique, Université de Montréal, Succ. Centre-Ville, Montréal, Québec, H3C 3J7, Canada}
\affiliation{Centre de recherche en astrophysique du Québec (CRAQ)}
\affiliation{Max-Planck-Institut f{\"u}r Astronomie, K{\"o}nigstuhl 17, D-69117 Heidelberg, Germany}

\begin{abstract}

Recent advances in machine learning algorithms have unlocked new insights in observational astronomy by allowing astronomers to probe new frontiers. In this article, we present a methodology to disentangle the intrinsic X-ray spectrum of galaxy clusters from the instrumental response function. Employing state-of-the-art modeling software and data mining techniques of the Chandra data archive, we construct a set of 100,000 mock Chandra spectra. We train a recurrent inference machine (RIM) to take in the instrumental response and mock observation and output the intrinsic X-ray spectrum. The RIM can recover the mock intrinsic spectrum below the 1-$\sigma$ error threshold; moreover, the RIM reconstruction of the mock observations are indistinguishable from the observations themselves. To further test the algorithm, we deconvolve extracted spectra from the central regions of the galaxy group NGC 1550, known to have a rich X-ray spectrum, and the massive galaxy clusters Abell 1795.  Despite the RIM reconstructions consistently remaining below the 1-$\sigma$ noise level, the recovered intrinsic spectra did not align with modeled expectations. This discrepancy is likely attributable to the RIM's method of implicitly encoding prior information within the neural network. This approach holds promise for unlocking new possibilities in accurate spectral reconstructions and advancing our understanding of  complex X-ray cosmic phenomena.

\end{abstract}

\keywords{}

\section{Introduction} \label{sec:intro}
Galaxy clusters harbor large reservoirs of hot gas ($\sim 10^7-10^8$~K), called the IntraCluster Medium (ICM), which account for the majority of baryonic matter in the cluster (e.g.,  \citealt{fabian_x-rays_2003}). 
This gas consists primarily of ionized hydrogen and helium but also contains numerous heavier elements (e.g.,  \citealt{mushotzky_x-ray_1984}; \citealt{mohr_properties_1999}; \citealt{loewenstein_chemical_2003}). Emission mechanisms such as thermal bremsstrahlung, bound-free atomic transitions, and the collisional excitation of hydrogen are responsible for the X-ray continuum (e.g.,  \citealt{markevitch_comparison_1997};  \citealt{ettori_coulomb_1998};  \citealt{sarazin_x-ray_1999};  \citealt{markevitch_temperature_1998}). The X-ray spectra of galaxy clusters also exhibit strong emission lines from heavier atoms (e.g., \citealt{fabian_observational_2012}; \citealt{sarazin_x-ray_1999}). In the soft X-ray regime ($0.5-2$~keV), several of these emission lines are present such as several Iron species (Fe XIX, Fe XVII, and Fe XVIII), Oxygen species (O VII), Nitrogen species (N VII), and Silicon species (Si XIII and Si XIV). The rest-wavelengths of these lines are contained in the atomic database for collisional plasma tailored to high-energy X-ray astrophysics, AtomDB (\citealt{foster_updated_2012}; this is primarily Chandra calibration data), and the calibration database, CalDB (\citealt{graessle_chandra_2006}). 
The underlying temperature of the ICM and the relative abundances of these metals have a substantial effect on the resulting X-ray spectrum (e.g., \citealt{fabian_x-rays_2003}; \citealt{markevitch_comparison_1997}; \citealt{mushotzky_x-ray_1984}); this enables the study of galaxy clusters through fitting plasma physics models to observed X-ray spectra.

The ICM was first observed over five decades ago; the first targets were the brightest and most nearby galaxy clusters: Perseus, M87, and Coma (\citealt{bradt_evidence_1967}; \citealt{gursky_strong_1971}; \citealt{gursky_x-ray_1973}). Since these first targets, X-ray observatories such as \textit{Uhuru} (\citealt{sarazin_x-ray_1986}), \textit{EXOSAT} (\citealt{giacconi_einstein_1979}), the \textit{EINSTEIN Observatory} (\citealt{forman_detection_1978}), \textit{XMM-Newton} (\citealt{jansen_xmm-newton_2001}; \citealt{wilman_fitting_1999}), and the \textit{Chandra} X-ray Observatory (\textit{CXO}) have revolutionized our understanding of the ICM (\citealt{forman_chandra_2002}; \citealt{forman_galaxy_2002}; \citealt{vikhlinin_evolution_2002}; \citealt{mazzotta_chandra_2001}). 

In particular, the \textit{CXO}, with an unprecedented spectral and spatial resolution has changed the field of X-ray astronomy (\citealt{weisskopf_chandra_2000}; \citealt{weisskopf_chandra_2005}; \citealt{weisskopf_making_2010}). This observatory uses a charge-coupled device to capture high-energy photons between approximately $0.3 - 10$~keV. Not only does it collect the spatial information of the incident photons, but it also registers their energy on the device. Consequently, we can construct a spectrum for each pixel on the CCD.

Due to its exquisite spatial resolution ($0.492\arcsec$ per pixel), the \textit{CXO} produces highly detailed, spatially resolved images of its targets. 
Precise spectroscopic measurements have allowed astronomers to map out temperature profiles and, in some cases, spatially resolved temperature maps (e.g., \citealt{bourdin_temperature_2004}; \citealt{pratt_temperature_2007}; \citealt{adam_mapping_2017}; \citealt{alden_galaxy_2019}). These maps, in conjunction with metallicity maps, allow for studying the energetic history of the ICM by tracing gradients in the thermodynamic parameters. Moreover, these observations, used in synergy with other observations at different wavelengths, have shed light on the role of the central supermassive black hole in regulating the ICM (\citealt{fabian_cosmic_2009}; \citealt{rafferty_feedback-regulated_2006}; \citealt{ruszkowski_supermassive_2019}). 

To extract these parameters, we fit theoretical models of the underlying astrophysics to the observed spectrum, considering any instrumental effects. One longstanding challenge in X-ray astronomy is that the observed X-ray spectrum, $S(E')$, is the result of a convolution between the source's intrinsic spectrum, $F(E)$, and the instrumental response, $R(E',E)$ as shown in the equation: 

\begin{equation}\label{eqn:spec}
    S(E') = \int_0^\infty R(E',E)F(E)dE, 
\end{equation}
where $E'$ denotes the discrete photon energies captured by the detector and $E$ is the measured energy.
Despite the simplicity of equation \ref{eqn:spec}, a direct deconvolution of the instrumental response function and model spectrum poses several issues (see \citealt{rhea_data_2021} for an in-depth discussion). Primarily, the instrumental response varies greatly as a function of position, and the detector energy space is limited and finite by design. 
Since the sampling of the detector space $E'$ is discrete, we can rewrite equation \ref{eqn:spec} as a matrix equation (\citealt{kaastra_optimal_2016}):
\begin{equation}\label{eqn:spec_mat}
    S_i = \sum_{j}R_{ij}F_j .
\end{equation}
We have replaced the instrumental response function with its matrix counterpart, $R_{ij}$. The first index, $i$, indicates the detector channel in $E'$-space, while the second index, $j$, denotes the emitted photon energy in $E$-space. In this formulation, $S_i$ represents the observed photon count rate in units of counts s$^{-1}$ for a given detector energy bin. $F_j$ is the model (also denoted as true) spectrum flux in units of counts m$^{-2}$ s$^{-1}$ in emitted energy bin $j$. We note that this discretization of equation \ref{eqn:spec} explicitly linearizes the equation. The ramifications of this choice are discussed in section \ref{sec:futureImprovements}.

As demonstrated in \cite{rhea_data_2021}, standard numerical methods, including different regularization methods, fail to deconvolve the true spectrum from the response matrix given the observed spectrum. Consequently, we began exploring a promising method known as the Recurrent Inference Machine (RIM) for deconvolution. This approach has shown remarkable success in deconvolving 2D radio images (\citealt{Morningstar_Analyzing_2018}; \citealt{Morningstar_Data_2019}), separating gravitational lens sources from their foreground counterparts (\citealt{adam_pixelated_2022}), and reconstructing MRI images (\citealt{lonning_recurrent_2019}).

In \cite{rhea_unraveling_2023}, we initially demonstrated the potential of this method for deconvolving X-ray spectra. Although astronomers have been able to fit the observed spectrum through a technique known as forward modeling, by which the theoretical model is convolved with the instrumental response and subsequently fitted, there are several benefits to having access to the intrinsic spectrum. Once deconvolved, these spectra can be fit directly, circumventing the need for forward modeling. Moreover, they can be used as inputs for machine learning algorithms such as a convolutional neural network to extract point estimates of the underlying thermodynamic parameters. 
This methodology can also study the line-by-line calibration of the \textit{CXO} by extracting the intrinsic spectrum of a non-changing calibration target where the intrinsic spectrum is not expected to evolve.

Therefore, we develop a new algorithm in this paper that can successfully disentangle the intrinsic X-ray spectrum from the instrumental response function. Throughout this paper, we limit the case to the X-ray spectra of galaxy clusters but stress that the method we develop could be applied to all fields of X-ray astronomy. In $\S$\ref{sec:methods}, we present the RIM and convolutional neural network used throughout this paper and the synthetic and natural X-ray observations used. We then discuss the results of using our RIM to deconvolve X-ray spectra in $\S$\ref{sec:results}. We compare fits using the deconvolved spectra with traditional methods, explore the reasoning behind the deconvolution, and explore failure modes in $\S$\ref{sec:Discussion}.
Finally, we conclude our paper in $\S$\ref{sec:conclusions}. We assume a Hubble constant of $H_0$=69.6, an energy density of matter, $\Omega_M$=0.286, an energy density of the vacuum, $\Omega_{vac}$=0.714, and a flat curvature.

\section{Methods and Observations}\label{sec:methods}
This paper aims to show that RIMs can successfully deconvolve the X-ray spectrum from the \textit{Chandra}  instrumental response function. Below, we outline in detail the machine and how it was trained.

\subsection{Recurrent Inference Machine}\label{sec:RIMmethods}
The RIM is an extension of a recurrent neural network (RNN) in which the RNN is used to provide updates to the solution mimicking a standard Newton-Raphson update; they have been used to solve 2D inverse problems (e.g., \citealt{Putzky_recurrent_2017}; \citealt{Morningstar_Analyzing_2018}; \citealt{Morningstar_Data_2019}). We adopt this architecture to solve our one-dimensional inverse problem described in equation \ref{eqn:spec_mat}. 

\begin{figure}[ht!]
    \centering
    \includegraphics[width=0.495\textwidth]{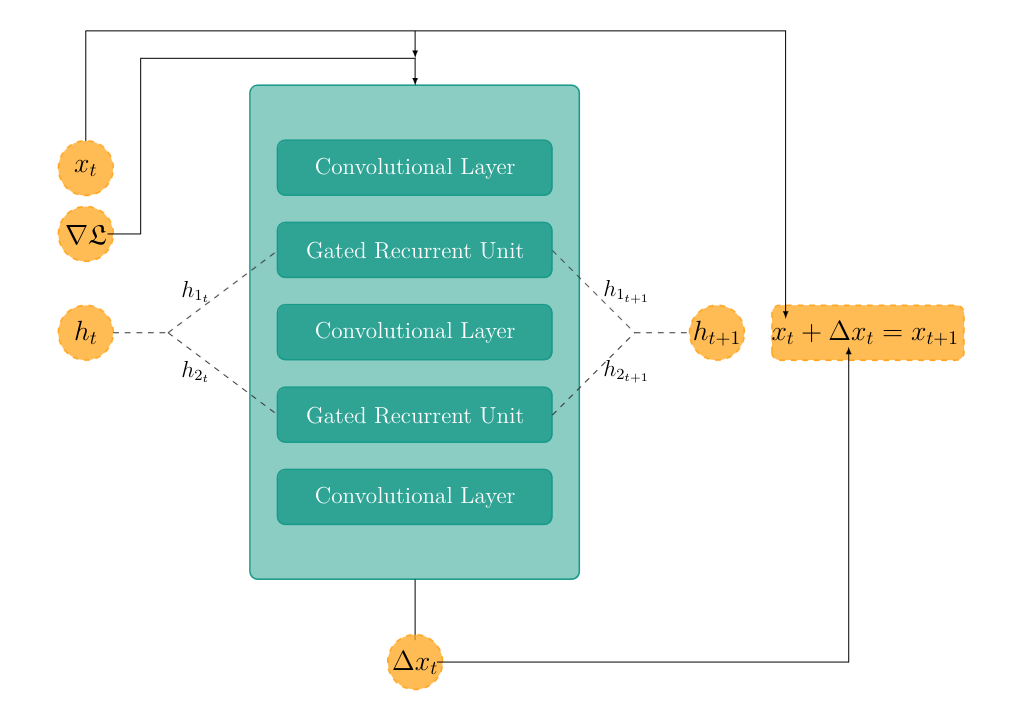}
    \caption{Schematic view of the RIM. Nodes in dark gray are treated as inputs to the RIM. Note that we separate the hidden layers before applying the RIM and combine them after applying the RIM. The teal box represents the RIM itself.}
    \label{fig:RIM}
\end{figure}

In general, the RIM solves the generalized linear problem
\begin{equation}
    \textbf{y} = A \textbf{x} + \textbf{n}
\end{equation}
by solving for \textbf{x}, the solution vector, recursively where \textbf{y} is the measurement vector, $A$ is some operation which manipulates the solution vector, and \textbf{n} is the noise vector.

At each step, the RIM uses the previous solution, $\mathbf{x}^{(t)}$, and the gradient of the log-likelihood function, $\grad \log p(\mathbf{y} \mid \mathbf{x}^{(t)})$, to solve for an update to the solution $\Delta \mathbf{x}^{(t)}$. The solution is updated at each step as
\begin{equation}\label{eqn:update}
    \mathbf{x}^{(t+1)} = \mathbf{x}^{(t)} + \Delta \mathbf{x}^{(t)} .
\end{equation}

The update of the solution can be viewed through the lense of a maximum a prosteriori (MAP) framework where
\begin{equation}\label{eqn:argmax}
    \hat{\textbf{x}}_{MAP} = \underset{\textbf{x}}{\text{argmax}} \big[ \log p(\textbf{y}|\textbf{x}) + \log p(\textbf{x}) \big]
\end{equation}

For this paper, we drop the prior term, $\log p(\textbf{x})$, following \cite{Morningstar_Analyzing_2018}.

More specifically, we can express the iterative solution, $\textbf{x}$ at time step $t+1$ as 

\begin{equation}\label{eq:recurrent series}
\begin{aligned}
    \textbf{x}^{(t)} &= \textbf{x}^{(t)} + g_\theta(\textbf{x}^{(t)}, \grad_{x}\log p(\textbf{y} \mid \textbf{x}), \mathbf{h}^{(t)}) \\
    \mathbf{h}^{(t+1)} &= g_\theta(\textbf{x}^{(t)}, \grad_{x}\log p(\textbf{y} \mid \textbf{x}), \mathbf{h}^{(t)})
\end{aligned}
\end{equation}

where $g_\theta$ is the neural network and $\theta$ represents the weights of the network, $p(\textbf{y}|\textbf{x})$ is the likelihood, and $\textbf{h}$ is the hidden vector used in the neural network.  

Figure \ref{fig:RIM} shows the standard architecture of a RIM (e.g., \citealt{Morningstar_Analyzing_2018}; \citealt{Morningstar_Data_2019}); the teal box represents the RIM itself. The methodology is as follows:
\begin{enumerate}
    \item Initialize the hidden vectors, $\mathbf{h}_1$ and $\mathbf{h}_2$, the initial solution, $\mathbf{x}^{(t=0)}$, and all weights of the gated recurrent units (GRUs)   and convolutional layers of the RIM.
    \item Calculate the gradient of the likelihood function $\grad \log p(\mathbf{y} \mid \mathbf{x}^{(t)})$.
    \item Pass the solution at the current time step, $\mathbf{x}^{(t)}$, the current value of the gradient of the likelihood function, $\grad \log p(\mathbf{y} \mid \mathbf{x}^{(t)})$, and the state vectors, $\mathbf{h}_1$ and $\mathbf{h}_2$ to the RIM.
    \item Forward pass through the neural network.
    \item Update the current value of the solution, $\mathbf{x}^{(t+1)}$ with the output of the RIM, $\Delta \mathbf{x}$, and the solution at the previous step, $\mathbf{x}^{(t)}$, following equation \ref{eqn:update}.
    \item Repeat steps 2-5 a predefined number of times.
\end{enumerate}

We note that the constituent layers of the RIM, the convolutional layers and GRUs function as normal and are updated through standard back-propagation following the Adam optimization scale (\citealt{kingma_adam_2017}). Following \cite{Morningstar_Data_2019}, we initialize $\mathbf{x}^{(t=0)}=0$ and each state vector, $\mathbf{h}^{(t)=0}_{1}$ and $\mathbf{h}^{(t=0)}_{2}$, as the zero vector. We defined the likelihood as 
\begin{equation}
   \log p(\mathbf{y} \mid \mathbf{x}^{(t)}) = -\frac{1}{2}(\mathbf{y}-A\mathbf{x}^{(t)})^TC^{-1}(\mathbf{y}-A\mathbf{x}^{(t)})
\end{equation}
where $C \in \mathbb{A}^{m \times m}$ is the covariance of the additive Gaussian noise of a particular observation. A Gaussian likelihood is ubiquitous in machine learning applications since the target distribution is often expected to be Gaussian.

Therefore, the gradient is written as
\begin{equation}\label{eqn:likelihood}
    \grad_{x^{(t)}} \log p(\mathbf{y} \mid \mathbf{x}^{(t)}) = (\mathbf{y} - A \mathbf{x}^{(t)})^TC^{-1}A
\end{equation}

By explicitly defining the physical model, we stress that our network is less of a black box since the physics of the problem is encoded in the likelihood. This allows us to incorporate the specific physics of the problem (i.e.,  the response matrix) in the updates. Furthermore, we apply an \textit{arcsinh} transformation function on the gradient of the likelihood function before passing it to the RIM to avoid exploding gradients.
We use a mean squared loss backpropagated through time function:
\begin{equation}
    \mathcal{L} = \frac{1}{T}\sum_{t=1}^T\sum_{i=1}^M (\hat{x}_i^{(t)} - x_i)^2
\end{equation}
where $\hat{\mathbf{x}}_i^{(t)}$ is the current best reconstruction at time $t$ and $M$ is the total number of spectral channels in the spectrum. Using this loss function, the RIM solution is considered at each time step.

In the case of the inverse problem investigated in this paper, we compute the loss function by comparing the result of our RIM with the actual spectrum. Thus, the RIM only sees the observed spectrum and the response matrix.
The layers of the RIM are the following:
\begin{enumerate}
    \item Convolutional Layer 1: 64 filters with a \texttt{tanh} activation function, stride equal 1, and padding equal 1
    \item GRU 1: 800 nodes with \texttt{sigmoid} activation function
    \item Convolutional Layer 2: 128 filters with a \texttt{tanh} activation function, stride equal 1, and padding equal 1
    \item GRU 2: 800 nodes with \texttt{sigmoid} activation function
    \item Convolutional Layer 3: 128 filters with a \texttt{linear} activation function, stride equal 1, and padding equal 1
\end{enumerate}
The convolutional filters all have a kernel size of 3.
We ran hyper-parameter tuning on the number of nodes in each GRU, filters and kernel size for each convolutional layer, and the batch size. We determined that eight convolutional filters and three kernel sizes were optimal. 

\subsection{Synthetic X-ray Data}\label{sec:syntheticDataMethodology}
To train the RIM to deconvolve the intrinsic spectrum from the response matrix, we need groups of response matrices, true intrinsic spectra (referred to as \textit{ground truth}) and the corresponding convolved observed spectra. Since we cannot access ground truth spectra for real observations, we rely on synthetic spectra to train our algorithm.
We create the synthetic spectra using the analysis software \texttt{SOXS}\footnote{https://hea-www.cfa.harvard.edu/soxs/}, which was created to generate high spectral resolution mock spectra for the Lynx observatory (e.g., \citealt{gaskin_lynx_2019}). To create the synthetic spectra, we first generate 100,000 mock spectra of the ICM using the \texttt{APEC} model -- a standard thermodynamic model that uses a combination of collisional ionization equilibrium (CIE) physics and non-equilibrium ionization (NEI) physics to model the emission of the ICM (\citealt{smith_collisional_2001}). CIE and NEI spectra are generated using AtomDB.  We also trained the model on 200,000 mock spectra and found no difference in the reconstructions. To generate the spectra, users are required to input the temperature of the gas, the metal abundance in solar units, the redshift of the gas, $z$, and the normalization of the model defined as 
\begin{equation}
    N = 10^{-14}\int \frac{n_e n_p}{4\pi(1+z)^2D_A^2} dV
\end{equation}
 where $n_e$ and $n_p$ are the densities of the electrons and protons, respectively, $D_A$ is the angular diameter distance to the gas, and $V$ is the volume of the gas. 
 We randomly select the gas temperature from a uniform distribution between 0.4 and 8 keV. This range is chosen because it coincides with the majority of gas temperatures in nearby clusters as revealed by \textit{Chandra} (e.g., \citealt{cavagnolo_entropy_2009}). The metallicity of the gas is randomly sampled from a uniform distribution between 0.2 $Z_\odot$ and 1.2 $Z_\odot$. Similarly, these values are standard in nearby galaxy clusters. We restrict the redshift to only nearby galaxy clusters below $z=0.02$ for this proof-of-concept. We set the normalization by randomly selecting a value from a uniform distribution between 0.1 and 10; we discuss the impact of this choice on the application to real spectra in section \ref{sec:realObsResults}. 
 We apply additional foreground absorption using the \texttt{tbabs} model (\citealt{wilms_absorption_2000}). We allow the column density, $N_{\rm{H}}$ to vary between 0.001$\times10^{20}$cm$^{-2}$ and 1 $\times10^{20}$cm$^{-2}$. The SNR was uniformly sampled between 5 and 100. These SNR values run the gambit of typical observations (e.g., \citealt{cavagnolo_entropy_2009})
 We provide the code to create the synthetic spectra in appendix \ref{app:soxs-data}.

A crucial free parameter governing synthetic data creation is the instrumental response. We limit our analysis to the \texttt{ACIS} instruments. Since the response matrix varies as a function of time and position on the chip, it is essential to sample the space of possible response matrices adequately. To do this, we mined the \textit{Chandra} archive of observations of well-studied galaxy clusters designated as such in \cite{mushotzky_x-ray_1984}.
We created 1000 response matrices by randomly sampling $10\arcsec$ regions from the chip where the cluster center was located.

\begin{figure*}
    \centering
    \includegraphics[width=0.98\textwidth]{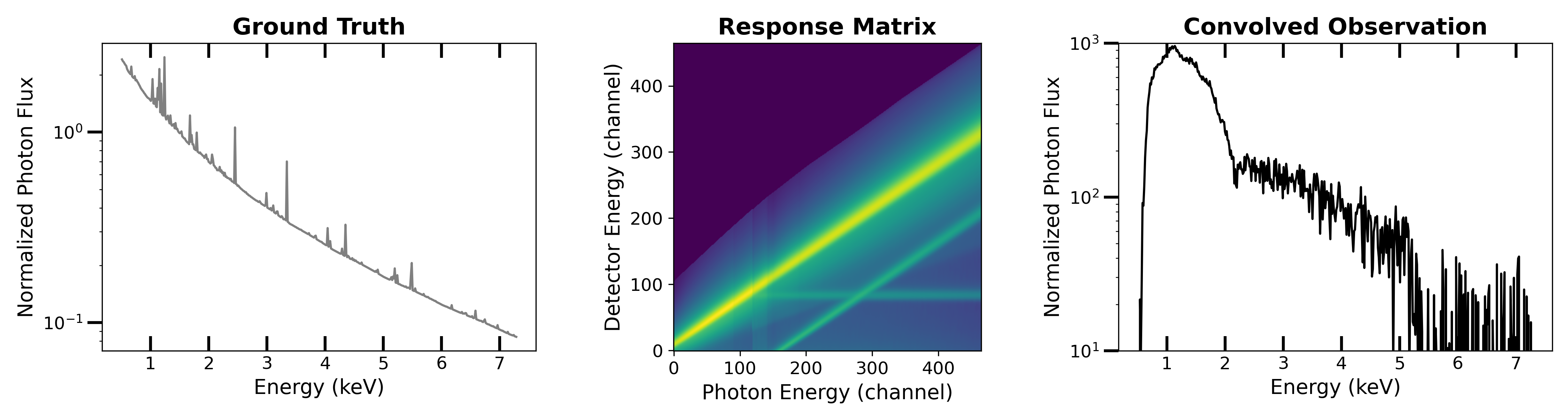}
    \caption{Schematic of the convolution applied by the ACIS instrument on \textit{Chandra}. This graphic demonstrates the response matrix's profound effect on the observed spectrum. The true intrinsic spectrum, the ground truth spectrum, is a typical emission spectrum from the ICM modeled using the \texttt{apec} model with a temperature of 2.0 keV and a metallicity of 0.3 Z$_\odot$. The response matrix was taken from a randomly chosen ObsID (2707); the matrix displayed is the log of the response matrix. The convolved observed spectrum was created following the methodology outlined in $\S$\ref{sec:syntheticDataMethodology}.}
    \label{fig:convolutionExample}
\end{figure*}

We create our training data on the fly using the forward model 
\begin{equation}
    S = R \otimes F + \mathfrak{N}
\end{equation}
where $\mathfrak{N}$ is a randomly sampled Gaussian distributon representing the noise. The sigma of the noisy distribution is randomly sampled, assuming a minimum and maximum signal to noise; we take the minimal value to be 10 and the maximum SNR to be 150. Below a SNR of 10, the observed X-ray data is dominated by the noise.
By building our training data, we can explore a vast parameter space describing the intrinsic emission model, response matrices, and noise configurations. We show an example of mock ground truth spectrum, response matrix, and convolved observed spectrum in Figure \ref{fig:convolutionExample}.

\subsection{Chandra X-ray Observations}\label{sec:ChandraObsMethods}
In this work, we apply our trained RIM to actual observations to validate its use. We select \href{http://cdsportal.u-strasbg.fr/?target=NGC\%201550}{NGC 1550} as our first test since it is a galaxy group that has a rich X-ray spectrum with a well-constrained temperature ($kT = 1.37\pm0.01$~keV) and metallicity ($Z\approx0.27 Z_\odot$; \citealt{sun_chandra_2009}; \citealt{kolokythas_evidence_2020}) at R$_{500}$. 
NGC 1550 is at a redshift z = 0.0124 (\citealt{sun_chandra_2009}). The \textit{CXO} has observed this target twice with the \texttt{ACIS-I} instrument. 

As our second test, we select the galaxy cluster \href{http://cdsportal.u-strasbg.fr/?target=Abell\%201795}{Abell 1795}. Contrary to NGC 1550, Abell 1795 has a considerably hotter ICM (kT$\approx$6 keV; \citealt{ehlert_very_2015}); due to this higher temperature, the intrinsic X-ray spectrum is expected to have fewer emission lines than NGC 1550; therefore, this cluster is a good juxtaposition. Abell 1795 is located at a redshift of z = 0.062476 (\citealt{oegerle_dynamics_2001}).
In Table~\ref{tab:chandraObsReal}, we detail these observations. The data were downloaded directly from the data archive, \href{https://cda.harvard.edu/chaser/}{\textit{chaser}}. All data were treated using \texttt{CIAO (v.14.5.1)}.

These objects were chosen since they represent two distinct points of the temperature scale in galaxy clusters and, thus, are represented by different spectral profiles. For NGC 1550, we use the two available observations to demonstrate how the deconvolution results change as a function of the observation. For Abell 1795, we select one random observation to highlight how the network responds to a different intrinsic spectrum. For each observation, the convolved observed spectrum is passed to the trained RIM, which, in turn, returns the deconvolved, intrinsic spectrum of the observation, referred to as the RIM Solution. 

\begin{table*}
    \centering
    \begin{tabular}{|c|c|c|c|}
       \hline
       Object & Chandra ObsID  & Date & Exposure Time (ks)  \\
       \hline \hline
       NGC 1550 & 3186 & 2002-01-08 & 9.65 \\
         \hline
       NGC 1550 & 3187 & 2002-01-08 & 9.99 \\ 
       \hline
       Abell 1795 & 5289 & 2004-01-18 & 14.95 \\
       \hline
    \end{tabular}
    \caption{Chandra ObsIDs used to test the RIM. We select two objects representing two extreme temperature scales in galaxy clusters, with NGC 1550 being a low temperature group and Abell 1795 being a massive hot galaxy cluster.}
    \label{tab:chandraObsReal}
\end{table*}

After downloading the data, we reduce and clean the level 1 event files using an in-house reduction pipeline located at \href{https://github.com/XtraAstronomy/AstronomyTools}{https://github.com/XtraAstronomy/AstronomyTools}. The pipeline first uses a CCD containing only background emission to construct a background light curves. We use the \texttt{CIAO} tool \texttt{lc\_sigma\_clip} to identify periods during the observation where the background differed from the 3-$\sigma$ level. These times were removed from the observations. We then destreak the event file and remove any bad pixels. The final processing is completed using \texttt{acis\_process\_events} with the \texttt{vfaint} option set to true since we are interested in the diffuse extended emission in the galaxy cluster. This entire process results in a level 2 event file for each ObsID. 
We extract the spectra within a radius of $R_{500}$\footnote{$R_{500}$ is a standard measure of the cluster's radius (e.g., \citealt{arnaud_universal_2010}). It is defined as $M_{500} = 500 (4\pi/3)\rho_c(z)R_{500}^3$ where $\rho_c(z)$ is the critical density of the universe at the cluster's redshift, and M$_{500}$ is the cluster's mass.} for the galaxy cluster using \texttt{specextract}, which creates the spectrum file and the two response files. These files are combined into a single response matrix using \texttt{rmfimg} with the argument \texttt{product=true}. We also extract a background spectrum from a region on the same CCD chip with no ICM emission.
We subtract this background spectrum, S$_{bkg}$, from the source spectrum, S$_{src}$, using the following equation:
\begin{equation*}
    \text{S}_{src} - \text{S}_{bkg} * \frac{\text{exp}_{src}*\text{bks}_{src}}{\text{exp}_{bkg} * \text{bks}_{bkg}}
\end{equation*}
where exp$_{src}$ and exp$_{bkg}$ are the exposure times of the source and background spectra, respectively, while bks$_{src}$ and bks$_{bkg}$ are the source and background backscales.

\section{Results}\label{sec:results}
 \subsection{Synthetic X-ray Data}\label{sec:SyntheticDataResults}
We run the RIM described in $\S$ \ref{sec:RIMmethods}  on 50,000 synthetic data created following the methodology outlined in $\S$\ref{sec:syntheticDataMethodology}. We assign 90\% 
The RIM was trained for 10 timesteps.
We remind the reader that the network's output is the deconvolved spectrum (also denoted as the RIM Solution) for each time step.
We applied a batch size of 64 and trained the network for 500 epochs. 
We tested different batch sizes and timesteps but found no effect on the final results.
\begin{figure*}
    \centering
    \includegraphics[width=1.05\textwidth]{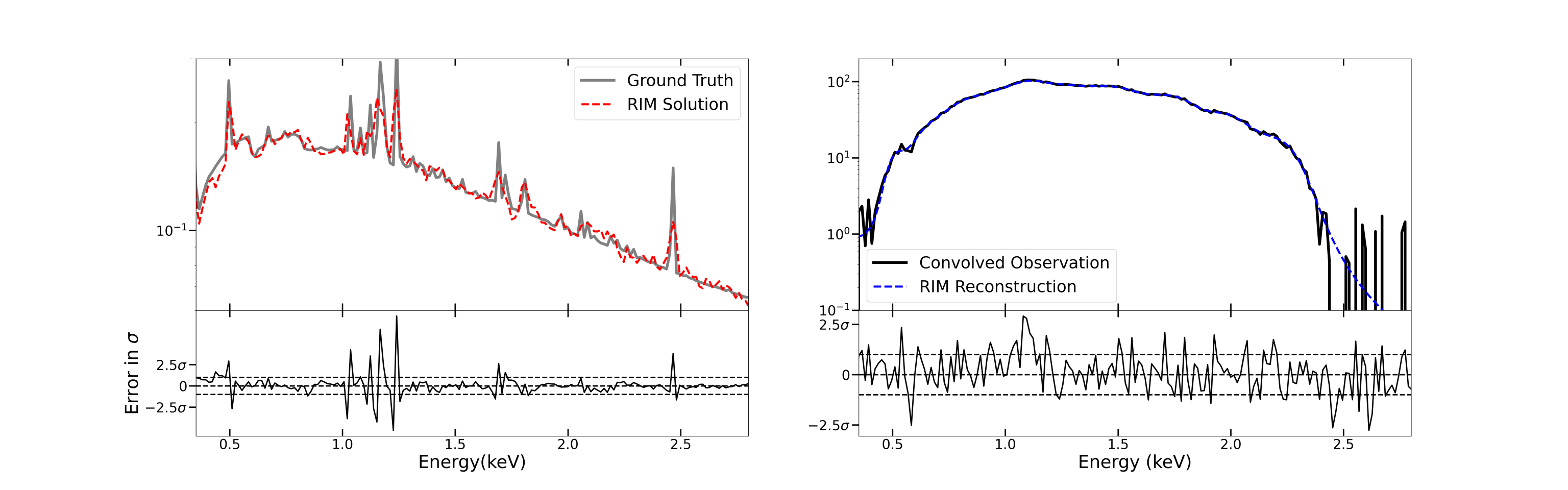}
    \caption{The RIM applied to synthetic X-ray spectra. In this figure, we show the results of the RIM deconvolution process on a randomly selected spectrum from the test set. We compare the ground truth in the upper left panel versus the RIM solution. In the upper right panel, we compare the convolved observed spectrum with the RIM reconstruction created by passing the RIM solution through the forward model. The bottom left panel shows the error between the ground truth and the RIM solution in units of $\sigma$. In the right bottom panel, we show the error between the convolved observed spectrum and the RIM reconstruction in units of $\sigma$.}
    \label{fig:syntheticObs}
\end{figure*}

\begin{figure*}
    \centering
    \includegraphics[width=1.05\textwidth]{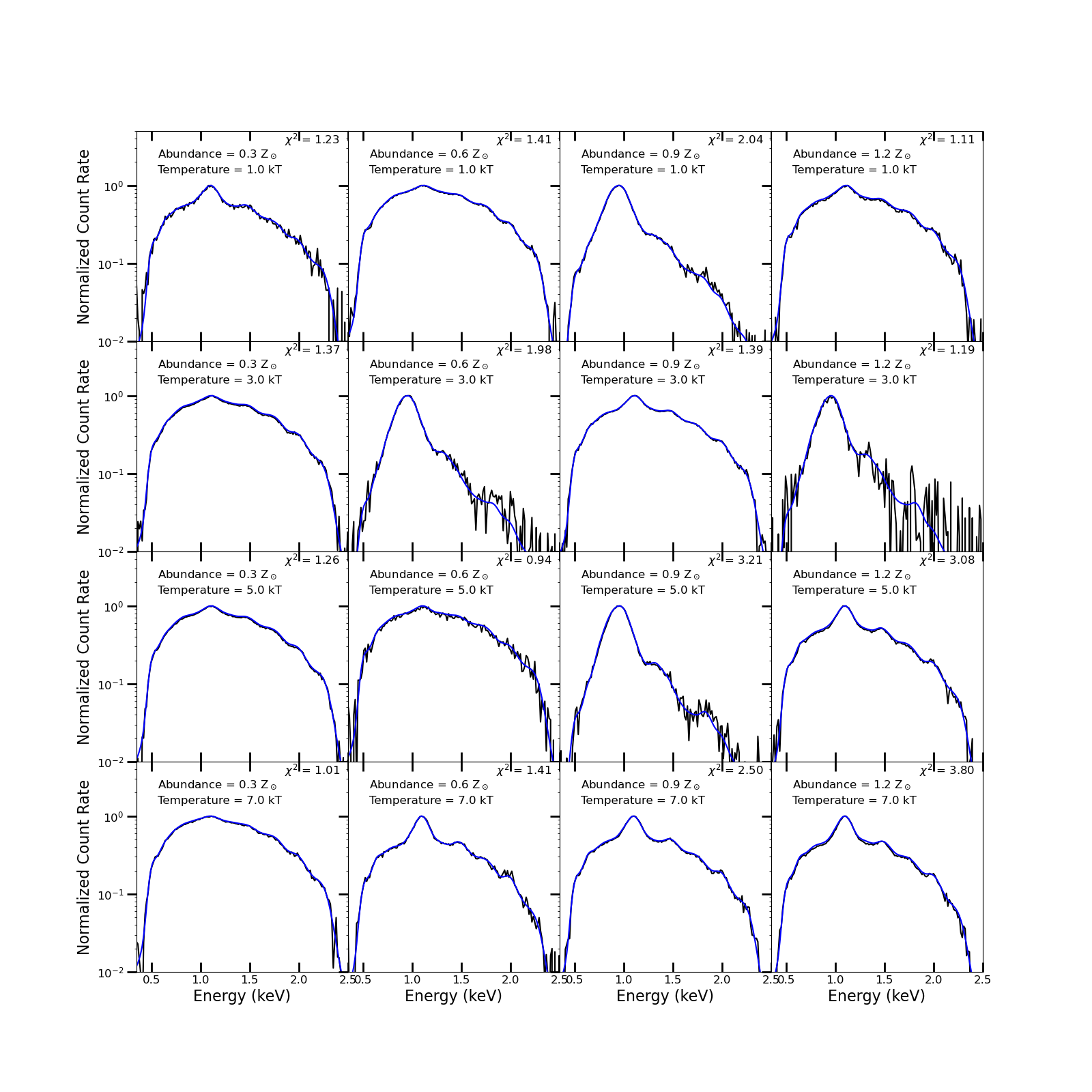}
    \caption{Montage of 16 randomly selected mock observed spectra (black) and the RIM reconstructions (blue) for a variety of temperatures and metalicities. Included in each subfigure is the reduced chi squared value. The temperature increases from top to bottom and the metalicity increases from left to right.}
\end{figure*}

In Figure~\ref{fig:syntheticObs}, we show the true intrinsic spectrum, i.e. the ground truth, vs. the RIM solution (top left), the convolved observed spectrum vs. the RIM reconstruction (top right), the error in the RIM solution as compared to the ground truth (bottom left), and the error in the RIM reconstruction as compared to the convolved observed spectrum (bottom right) for a randomly selected test spectrum. The RIM solution is the direct output of the RIM, while the RIM reconstruction is the result of passing the RIM solution through the forward model (equation \ref{eqn:spec}). The left-hand plot demonstrates that the RIM solution agrees with the literature model under 1-$\sigma$. The most significant errors occur in the region between 0.5 and 1.0 keV where the strong sulfur, silicon, and lithium emission lines are present (e.g., \citealt{markevitch_comparison_1997}; \citealt{mushotzky_x-ray_1984}; \citealt{sarazin_x-ray_1999}). While the errors remain below the 1-$\sigma$ level, we note that the RIM misses small-scale features while globally reconstructing the spectrum; this has been noted for previous use cases of the RIM (e.g., \citealt{adam_posterior_2022}).
The right-hand plot demonstrates that the RIM reconstruction is indistinguishable from the mock observation to a 1-$\sigma$ level except for the same region between 0.5 and 1.0 keV. This indicates that the RIM is learning the noiseless intrinsic spectrum to an accuracy below 1-$\sigma$ except in this region. 

Since this region is physically essential, we attempted several changes to the hyper-parameters of the RIM, including a dynamic learning rate scheduler and training for more extended periods. However, none of these changes improved the results of the RIM in this critical section. We explore the ramifications of this in section \ref{sec:limitations}.

\subsection{Chandra X-ray Observations}\label{sec:realObsResults}
We use both the extracted spectra and their associate response matrices to deconvolve the R$_{500}$ spectra created in section \ref{sec:ChandraObsMethods} for NGC 1550 and Abell 1795. 

\begin{figure*}
    \centering
    \includegraphics[width=0.98\textwidth]{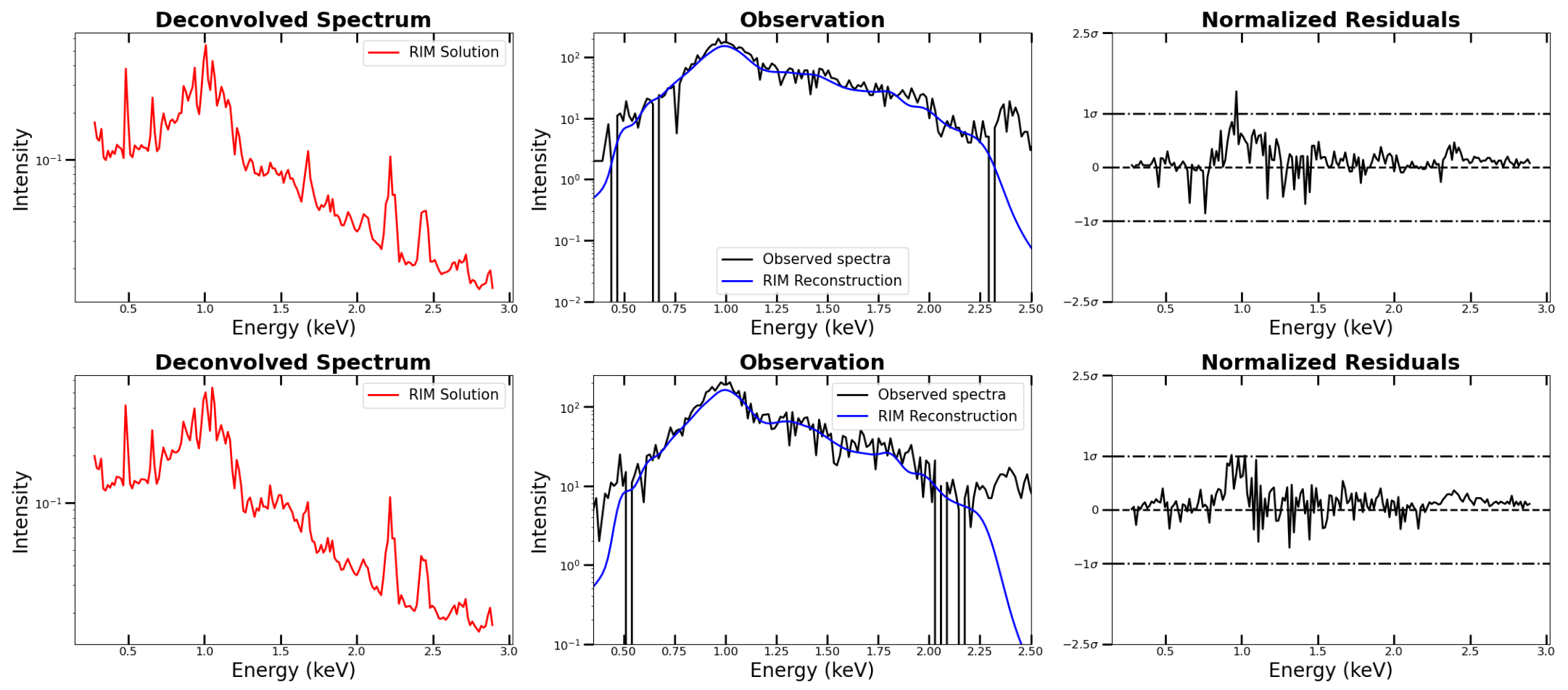}
    \caption{Results of the RIM on the galaxy group NGC 1550. In the left panels, we show the RIM solution (i.e. the result of the deconvolution process) for ObsID 3186 and 3187, top and bottom, respectively (red). The middle panel shows the observed spectrum after background subtraction (black) and the RIM reconstruction (blue). The RIM reconstruction results from passing the RIM solution through the forward model developed in equation \ref{eqn:spec_mat}. The right panel shows the  residual between the observed spectrum and the RIM reconstruction normalized to the noise level of the observation.}
    \label{fig:realObs}
\end{figure*}

To normalize the data for NGC 1550, the spectra are multiplied by the observation's exposure times, 9,650 seconds and 9,990 seconds, respectively, for ObsID 3186 and ObsID 3187. 
In doing so, we ensure that the data fed to the network resembles the magnitude of the synthetic data.
The noise level of the observation is estimated as the standard deviation of the intensity over the entire bandpass.
In Figure \ref{fig:realObs}, we show the results of the RIM deconvolution for ObsID 3186 (top) and ObsID 3187 (bottom). In the left panel, we show the direct result of the RIM deconvolution (red), i.e. the RIM solution. In the center panel, we display the observed spectrum (black) and the RIM reconstruction (blue). We show the normalized residuals between the observed spectrum and the RIM reconstruction in the right panels. These results demonstrate that the RIM reconstruction matches the observed spectra below the 1-$\sigma$ level except for in the busiest area of the spectrum around 1 keV. We explore the limitations of this reconstruction in $\S$ \ref{sec:limitations}.

\begin{figure*}
    \centering
    \includegraphics[width=0.96\textwidth]{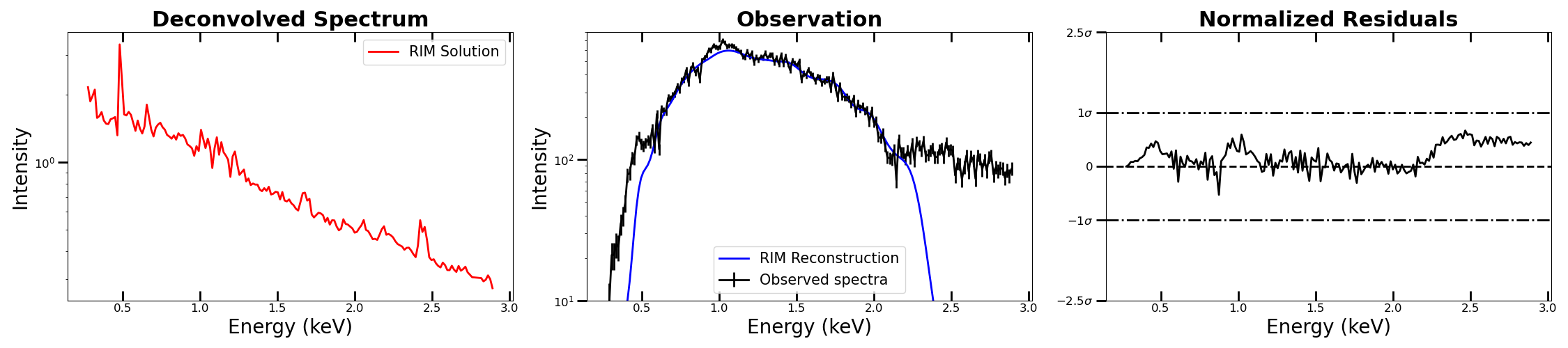}
    \caption{Results of the RIM on the massive galaxy cluster Abell 1795. In the left panels, we show the RIM solution (i.e. the result of the deconvolution process) for ObsID 5289 (red). The middle panel shows the observed spectrum after background subtraction (black) and the RIM reconstruction (blue). The RIM reconstruction results from passing the RIM solution through the forward model developed in equation \ref{eqn:spec_mat}. The right panel shows the residual between the observed spectrum and the RIM reconstruction normalized to the noise level of the observation.}
    \label{fig:realObs5289}
\end{figure*}

In Figure \ref{fig:realObs5289}, we show the results of the spectral deconvolution of ObsID 5289 (Abell 1795). Here, we used an exposure time of 14,950 seconds to normalize the data. Since Abell 1795 has a high temperature, the complex of strong emission lines around 1.0 keV is not as prominent as with NGC 1550. Figure \ref{fig:realObs5289} demonstrates similarly that the RIM reconstruction matches the observed spectrum below the 1-$\sigma$ level in all spectrum regions.

\section{Discussion}\label{sec:Discussion}

\subsection{Reasoning Behind Deconvolution}
Although the community has been working with the convolved X-ray spectra for the last several decades, deconvolution of X-ray spectra promises many benefits. Using deconvolved spectra, it will be possible to directly stack X-ray observations of the same target to improve the signal-to-noise. In doing so, we will no longer need to rely on simultaneous fitting of multiple observations, which is costly. 

Moreover, the deconvolved spectra can be passed to machine learning algorithms capable of extracting the underlying model parameters (i.e. the temperature and metallicity) without worrying about the response matrices. The benefit of this methodology has been shown extensively in the literature (e.g., \citealt{fabbro_application_2018}; \citealt{rhea_machine_2020}; \citealt{rhea_machine_2021}).

X-ray time variability has been the focus of many studies (e.x. \citealt{mushotzky_x-ray_1993};  \citealt{giannios_spectra_2004}; \citealt{van_der_klis_rapid_2006}); these studies tend to focus on either photometric measurements, or, in the case when they are studying the spectra, model fits to convolved spectra. Our proposed methodology to deconvolve X-ray spectra could be used to study the variability of X-ray sources in time. Typically, changes to the spectrum due to time variability are nearly indistinguishable due to the convolution. However, if we have access to the deconvolved spectrum, the evolution of the spectrum will be clear without the use of additional modeling. 

The RIM offers new possibilities for verifying the consistency of the CXO calibration over time. Indeed, contamination layers on Chandra's optical blocking filters cause the instrumental response of its detectors to evolve, leading to changes in the spectra of stable astrophysical sources observed across different cycles. These sources, which should have consistent spectra over time, are used as calibration sources to update the response matrix models, which are then used for data processing and spectral analysis. If the RMF calibration is accurate, the RIM-deconvolved spectra of these non-changing calibration targets across different cycles should be identical. If not, the RIM can be used to adjust the RMF by deconvolving recent spectra and comparing them to a reference spectrum from the initial cycle, which is assumed to be accurately calibrated. The coefficients of the response matrix are iteratively modified until the deconvolved spectra match the reference, minimizing the residual and resulting in a recalibrated matrix.

\subsection{Limitations of the Model}\label{sec:limitations}
In Section \ref{sec:SyntheticDataResults}, we demonstrated that the RIM reconstructions match the synthetic ground truth spectra well below the noise level. However, we do not have access to the ground truth for the real data. Therefore, we instead compare the RIM solution of the spectrum predicted from the best-fit model in the literature. 

\begin{figure}
    \centering
    \includegraphics[width=0.495\textwidth]{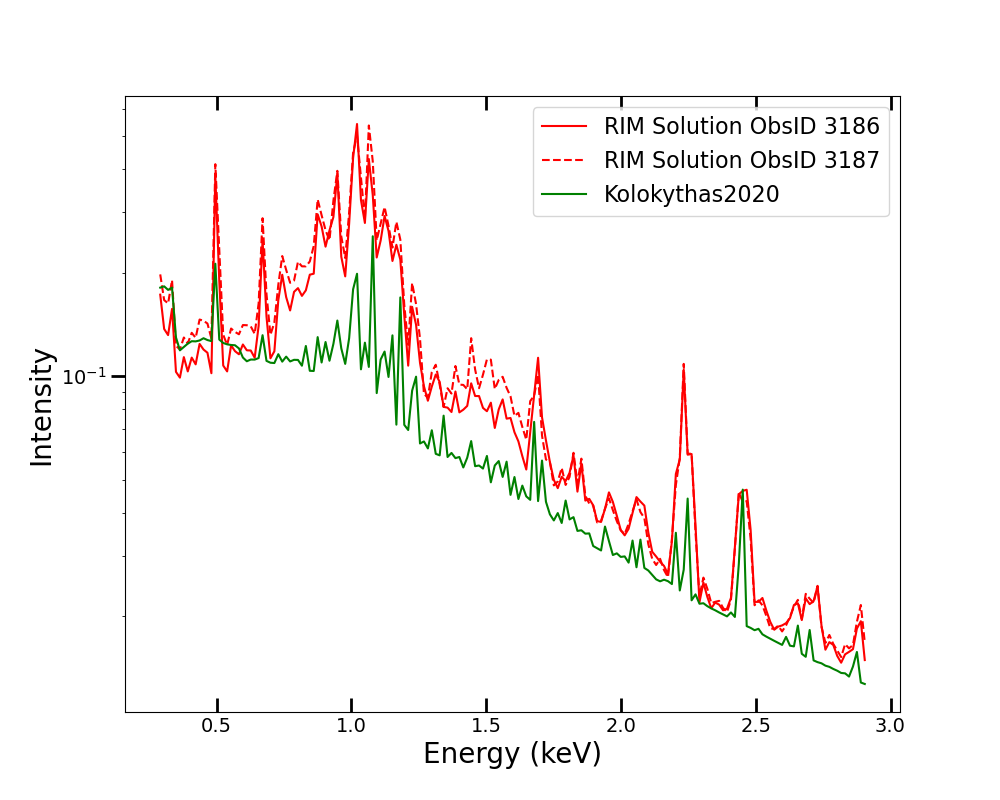}
    \caption{In this figure we compare the RIM solution for both ObIDs (ObsID 3186 and ObsID 3187 in red) and the model taken from literature values (green) for NGC 1550. The later was obtained by estimating the thermodynamic properties of the cluster from the convolved observed spectrum in Kolokythas2020 and then creating a mock spectrum using in \texttt{SOXS}.}
    \label{fig:modelComparison}
\end{figure}

Here is the corrected version while keeping all LaTeX commands and symbols intact:

More specifically, in the case of NGC 1550, we use the best-fit model estimated in \citet{kolokythas_evidence_2020} and refer to this as \textit{kolokythas2020} from now on. Here, the authors fitted the convolved observed spectrum and estimated that the ICM can be modeled by an absorbed \texttt{APEC} model, where the metallicity is set to 0.27~Z$_\odot$ and the temperature is set 1.38~keV. The only parameter we modify is the normalization parameter to match the deconvolved spectra, i.e., the RIM solution. In Figure \ref{fig:modelComparison}, we show the comparison between the model predicted from \textit{kolokythas2020} (dashed line) and the RIM solution from ObsID 3186 and ObsID 3187 (solid line). We stress that since we are deconvolving a real observation, there is no ground truth to compare the results to, but rather a fit that implicitly carries all the assumptions that go into the standard fitting procedure.


First, Figure \ref{fig:modelComparison} shows that, while the deconvolution of ObsID 3186 and ObsID 3187 return similar intrinsic spectra, there are discrepencies between the two deconvolutions. Galaxy clusters change on the timescales of tens of millions of years, therefore it is unexpected for the intrinsic spectrum to change. We consider the driving force between the differences in deconvolution to be the considerable singularity of the response matrices which differ from one observation to the other even though the RIM has converged and shows excellent results on realistic synthetic data. This result highlights the intrinsic difficulties in deconvolving the X-ray spectrum from the response matrix. 

More importantly, the deconvolved spectra do not match the expected model spectrum using the \textit{Kolokythas2020} model. The \textit{Kolokythas2020} model differs from the RIM solution in three main ways: the intensity of the powerlaw is slightly diminished, the intensity of the iron features around 0.8 - 1.2 keV are subdued, and the peak of the emission line around 2.2 keV is lower. This result can be interpreted as either the RIM solution, although relatively stable over different observations, is incorrectly capturing the intrinsic emission due to the inherent complexities of the inversion process, or the \textit{Kolokythas2020} model does not accurately represent the intrinsic emission. It is impossible to distinguish between the two scenarios without knowing the true result, which would require a higher resolution observation.

\begin{figure}
    \centering
    \includegraphics[width=0.495\textwidth]{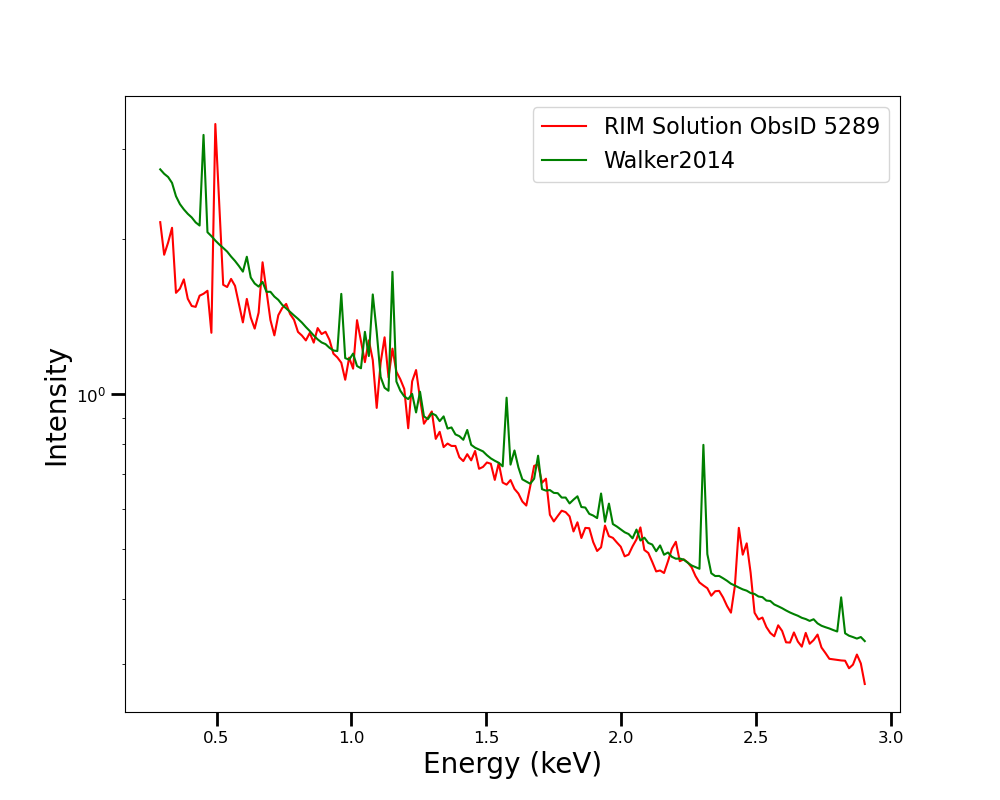}
    \caption{Same as in Fig. \ref{fig:modelComparison}, but for Abell 1795. In this figure we compare the RIM solution for ObsID 5289 (red) and the model taken from literature values (green) taken from Walker2014.}
    \label{fig:modelComparison-5289}
\end{figure}

Figure \ref{fig:modelComparison-5289} shows the equivalent RIM solution for ObsID 5289 and the expected model spectrum of Abell 1795. We calculate the expected model spectrum by assuming a temperature of kT $\approx5$ keV and a metallicity of Z $\approx0.5$ Z$_\odot$; these values were approximated from \cite{ettori_deep_2002} and \cite{walker_exploring_2014}; we therefore refer to this as the \textit{Walker2014} model from now on. The galactic absorption was set to 1.2  $\times 10^{20}$ cm$^{-2}$ following measurements by \cite{kalberla_leidenargentinebonn_2005}.
While the RIM solution more closely matches the literature model, there are still discrepencies between the two in the location and amplitude of the emission lines.

Despite these challenges with actual X-ray observations, the RIM performs remarkably well on synthetic data, consistently achieving highly accurate reconstructions that closely match the ground truth. This demonstrates the robustness of the method in controlled scenarios. However, further investigation is required to address the complexities of applying the RIM to actual observed data, which may involve refining the model and accounting for observational uncertainties. We discuss this further in the next section. Additionally, our results suggest that the thermodynamic parameters estimated from the convolved observed spectra may not fully capture the intrinsic properties of the cluster, potentially requiring a more nuanced approach.

\subsection{Future Improvements}\label{sec:futureImprovements}
Given the inherent challenges of solving highly ill-posed inverse problems, here we investigate more state-of-the-art approaches that could improve the accuracy of the RIM in future applications.

\cite{adam_pixelated_2022} applied the RIM to strongly gravitationally lensed systems, developing a method to simultaneously reconstruct undistorted images of the background source and the lens mass density distribution as pixelated maps. While successful, the authors showed that fine-tuning the RIM's objective function was necessary to improve reconstructions when fine structure was present in the data. Although this approach did not entirely eliminate the issue, it significantly improved the results. This methodology could potentially be adapted to X-ray spectra; however, since our current reconstructions are already at the noise level of the observations, this technique may have limited impact in our specific case.

The root of the problem lies in the way the RIM implicitly encodes the prior inside the neural network trained using the maximum likelihood estimate. Therefore, the RIM will produce blurred reconstructions because it returns an average spectrum corresponding to an observation.

Recent advances in  diffusion models have shown marked improvement over other techniques such as the RIM (e.g., \citealt{adam_posterior_2022}; \citealt{dia_bayesian_2023}; \citealt{adam_echoes_2023}). Diffusion-based techniques achieve better results by accurately encoding the priors using score-based models that implicitly do not require the prior to be learned. 

Therefore, a future paper will use diffusion models to recover the intrinsic spectrum. 
Furthermore, since diffusion models are anchored in Bayesian statistics, this will also give us uncertainties on the reconstructions, which is crucial for comparing the deconvolution results with contemporary models bound by literature results.

Another potential improvement arises from our assumption when transitioning from equation \ref{eqn:spec} to equation \ref{eqn:spec_mat}, where we linearized each term in the integral equation. The linearization of equation \ref{eqn:spec} implictly assumes that the response matrix behaves in a linear manner from one bin in energy space to another. This is potentially true for high resolution spectra such as those obtained with the X-Ray Imaging and Spectroscopy Mission (XRISM), but it is likely not the case for \textit{Chandra} ACIS-I or ACIS-S observations. If this is the case, then additional terms need to be added to equation \ref{eqn:spec_mat} that capture the nonlinear nature of the initial spectral equation for the two equations to be truly equivalent. 

Therefore, in the case of new and upcoming high spectral resolution missions such as XRISM, new Athena, or LEM, we expect the linearization condition to hold since it requires a small bin size in energy space (\citealt{kaastra_optimal_2016}). The RIM provides an accurate and efficient way in obtaining the actual intrinsic spectra of X-ray sources for these observatories. 

Furthermore, high spectral resolution missions will, by design, produce high resolution spectra where the response matrix does not cause strong degenerecies in the physical parameter space employing mixing emission lines. In future works, we will explore additional terms in the linearization of equation \ref{eqn:spec} and apply the RIM to future observatories.

Another potential challenge in recovering the intrinsic spectrum of real observations is that the ICM has a variety of temperatures but is rather fully characterized by a distribution of temperatures. While this poses a problem for the RIM which outputs a single expected spectrum, the diffusion-based model can capture this distribution in the posterior it will provide.

\section{Conclusions}\label{sec:conclusions}
In this paper, we explored using a machine learning algorithm, the RIM, to deconvolve the intrinsic X-ray spectrum of an emitting source from the instrumental response function observed using data from the \textit{Chandra} X-ray Observatory. 

We demonstrate that the RIM effectively reconstructs the global features in modeled ICM spectra below a 1-$sigma$ noise level. Moreover, the RIM-reconstructed spectra match the realistic synthetic observations closely. 

Furthermore, we apply our trained RIM to actual observations of the NGC 1550 and Abell 1795 galaxy clusters. The RIM, again, achieves a reconstruction error under 1-$\sigma$. While these results represent a significant improvement over previous attempts to deconvolve X-ray spectra from the instrumental response, they highlight the method's inability to reconstruct local changes in the reconstructed spectra. In the case of using the deconvolved spectra for science, the local features are crucial since they encode physical information about the emitting gas such as the underlying temperature or metallicity. 

However, this assumes that the expected models are correct, which implicitly asserts that the literature values of thermodynamic parameters are accurate. While not unreasonable, these results present a different intrinsic spectrum of Abell 1795 and NGC 1550; this underscores the need to validate that the RIM results are physically consistent.

Future works will use carefully constructed priors to augment the RIM and score-based models for deconvolving X-ray spectra.

\section{Software and third party data repository citations} \label{sec:cite}


%

\vspace{5mm}
\facilities{CXO}


\software{astropy }



\appendix

\section{Chandra Observations for Response Matrices}\label{app:responseMatrices}
In this section, we enumerate the \textit{Chandra} observations used to construct the response matrix catalog.

\subsection{SOXS Data Creation Commands}\label{app:soxs-data}
This section provides the code used to create the sythetic spectra.

\begin{python}
    from soxs import ApecGenerator, RedistributionMatrixFile
    rmf = RedistributionMatrixFile("region1.rmf")  # THIS IS ONLY USED TO GET THE RANGE AND SPACING
    emin = rmf.elo[0]
    emax = rmf.ehi[-1]
    nbins = rmf.n_e
    binscale = "linear"
    agen = ApecGenerator(emin, emax, nbins, binscale=binscale)

    num_spec = 100000
    true_ys = []
    for i in tqdm(range(num_spec)):
        # Randomly select xspec parameters
        redshift = random.uniform(0.0, 0.8)  # redshift
        temp = random.uniform(0.4, 8.0)  # temperature in keV
        abundance = random.uniform(0.2, 1.2)  # Metallicity abundance in Z_solar
        kT = (temp, "keV")
        norm = random.uniform(0.1, 1)
        spec = agen.get_spectrum(kT, abundance, redshift, norm)
        n_H = random.uniform(0.001, 0.01) # Foreground Galactic Absorption
        spec.apply_foreground_absorption(n_H, model="tbabs")
        true_ys.append(spec.flux)

\end{python}

\bibliography{CNN-Xray}{}
\bibliographystyle{aasjournal}



\end{document}